\documentclass[twoside,12pt]{article}
\usepackage{epsfig}
\usepackage{amssymb}
\title{Theoretical remark on the superconductivity of metals\footnote{from ``Gedenkboek aangeb. aan H. Kamerlingh Onnes, eaz. Leiden, E. IJdo, 1922, pp. 435'' /
translated with courtesy of the Kamerlingh Onnes Laboratory, Leiden - Institute of Physics, Leiden University}
}
\author{A. Einstein \\
translated by Bjoern S. Schmekel (Cornell University)\footnote{new address effective September 1, 2005: 
Department of Astronomy, University of California, Berkeley, California 94720, e-mail: {\rm schmekel -at- berkeley.edu}} }

\date{}

\begin{document}
\maketitle

The theoretical oriented scientist cannot be envied, because
nature, i.e. the experiment, is a relentless and not very friendly 
judge of his work. In the best case scenario it only says ``maybe'' to a theory,
but never ``yes'' and in most cases ``no''. If an experiment agrees with theory
it means ``perhaps'' for the latter. If it does not agree it means ``no''.
Almost any theory will experience a ``no'' at one point in time - most theories
very soon after they have been developed. In this paper we want to focus on
the fate of theories concerning metallic conductivity and on the revolutionary influence
which the discovery of superconductivity must have on our ideas of metallic conductivity.

After it had been recognized that negative electricity is caused by subatomic carriers of particular
mass and charge (electrons), there were good reasons to believe that metallic conductivity 
rests on the motion of electrons. Furthermore, the fact that heat is conducted much better
by metals than by non-metals as well as the Wiedemann-Franz law about the substance-independence 
of the ratio of electric and thermal conductivity of pure metals (at room temperature) led to attribute the thermal 
conductivity to electrons as well. Under these circumstances there were
reasons for an electron-based theory of metals similar to the kinetic gas
theory (Riecke, Drude, H. A. Lorentz). In this theory free electron motion
is assumed which resembles gas molecules with thermal mean kinetic energy
$3/2~kT$ neglecting collisions with metal atoms. This theory was remarkably successful in the sense that it could predict the coefficient in the
Wiedemann-Franz law from the ratio of mechanical and electric mass of the electron with remarkable precision. It also explained qualitatively
the appearance of thermo-electricity, the Hall effect etc.
No matter how the theory of electron conductivity may develop in the future one main
aspect of this theory may remain valid for good, namely the hypothesis that
electric conductivity is based on the motion of electrons.

The Drude formula for the specific resistance $\omega$ of metals is

\begin{eqnarray}
\omega = \frac{2m}{\epsilon^2} \frac{u}{nl}    
\label{eq:1}
\end{eqnarray}
where $m$ is the mass, $\epsilon$ the charge of the electron, $u$ the average velocity, $n$ the volume density and $l$ is the free path length of the electron.
Unfortunately, there are three unknown temperature functions $u$, $n$, $l$ in the
theory; one of them ($u$) is related to the absolute temperature according to
\begin{eqnarray}
3mu^2 = kT ~; 
\label{eq:2}                                 
\end{eqnarray}
n has to be small compared to the mean density of atoms ensuring 
that the electrons do not contribute to the specific heat of the metal.
To what extent is Eq.~(\ref{eq:1}) suitable to explain the dependence of the
specific resistance on the temperature?
You will get into difficulties. According to Eq.~(\ref{eq:2}) u should be proportional
to $\sqrt{T}$. One does not expect a significant dependence of the path length $l$ on the temperature. One would have to expect a rapid increase in the number
n of electric dissociated atoms as the temperature increases, because 
the dissociation of a weakly dissociated substance grows rapidly with $T$.
One ought to think that the resistance of a pure metal decreases as the temperature increases. However, this is not the case since it is well known 
that the resistance is proportional to $T$ at high temperatures. 
Considering this characteristic fact due to Eq.~(\ref{eq:1}) one would have to
look at the hypotheses: the number n of free electrons is independent of
temperature; the free path length of the electrons is inversely proportional 
to the root of the energy content of the metal.
With Eq.~(\ref{eq:1}) having been modified like this Kamerlingh Onnes was able to
describe the properties of metals in the non-superconducting state with
amazing precision. The hypothesis of the dependence of the path length on
the thermal agitation is not too peculiar; one could imagine that the electron
in an agitation-free metal moves as if in empty space, but inhomogeneities
due to thermal excitations provide electric fields which deflect the electrons. The hypothesis of the temperature-independence of $n$ is questionable, though.
Also, it may be hard to account for the relation between $l$ and heat content
in a quantitative manner. In any case the success of Kamerlingh Onnes's 
thoughts prove that the thermal agitation of metals (not of electrons)
is mainly the resistance-provisory moment. Only this may explain
why the resistance at high temperatures obeys the law
\begin{eqnarray}
\omega = \alpha (T - \theta)
\end{eqnarray}
and not the law\footnote{cf. e.g. Comm. No. 142a, Versl. Ak. Amsterdam, June 1914, Fig. 2 for Sn, Cu, Cd and Suppl. No. 34b, Report
third Int. Congr. Refr. Chicago, Fig. 5 for Hg}
\begin{eqnarray}
\omega = \alpha T
\end{eqnarray}
and why the resistance of non-superconducting metals
becomes independent of temperature at low temperatures. 
The curvature of the resistance curve at low temperatures is thereby indirectly related to quantum theory.
Actually, according to this notion the resistance of non-superconducting metals
should approach zero as the temperature decreases, but it is found 
that the resistance approaches a non-zero limit. Kamerlingh Onnes found
that the actual limit is influenced very strongly by small amounts of
impurities. Furthermore, these impurities shift the whole resistance
curve vertically, i.e. they cause an ``additive resistance'' such that
the resistance of the pure homogeneous metal could have zero resistance
as the limit. It should be mentioned that this remarkable fact is 
incompatible with the explanation provided by Eq.~(\ref{eq:1}) . It can be
shown easily that in the presence of opportunities for a collision 
a constant is added to $1/l$. However, this constant does not alter
the resistance by a temperature-independent amount, but rather by
an amount proportional to u (or u/n, respectively); under no circumstances
can u assumed to be temperature-independent, because otherwise the only
success of the theory, i.e. the explanation of the Wiedemann-Franz law,
would have to be sacrificed. For the same reason it may be hard to find
a theoretical explanation of the constant resistance of metals with impurities
at low temperatures. 
This outline shows that the thermal electron theory fails even for common
conductivity - not to mention superconductivity. On the other hand
it is conceivable that the Wiedemann-Franz law follows from a 
different theory which attributes electrical and thermal conductivity to
an electron-mechanism.

The breakdown of the theory became obvious after superconductivity of metals had been discovered. Kamerlingh Onnes made a convincing case that superconductivity
cannot be based on electrons with thermal agitation by showing that non-superconducting wires with a thin coating made of a superconducting material
become superconducting. The electrons in the coating would have to penetrate
into the non-superconductor and would lose their preferred mean motion
which causes the current. The system would have to be non-superconducting.

If one wanted to explain superconductivity by [the presence of] free electrons
one would have to view them as agitation-free such that the negative electricity
in the current-carrying superconductor has no other motion besides
the motion that makes up the electric current. Such a view is implausible
not only because of the Rutherford-Bohr theory, in which strong electric
fields exist in the interior or a body, but also because of the fact that
superconductivity is destroyed by moderate magnetic fields. Because the transverse forces due to the Lorentz force (Hall force) would be compensated
electrostatically at the surface by charge accumulation in such a way that
no effect of the magnetic field on the electrons would be expected.

It seems that electric conductivity has to be attributed to the peripheral
electrons of the atoms which move around the nucleus at high velocity.
Indeed, according to the Bohr theory it seems hardly conceivable that
the circulating peripheral high-energy electrons (e.g. the ones of mercury vapor)
lose a significant part of their velocity at condensation which is comparatively 
speaking not too constraining energetically. Given our present
knowledge it seems as if free electrons did not exist in metals. Then metallic
conductivity is caused by atoms exchanging their peripheral electrons.
If an atom received an electron from a neighbouring atom without giving an electron to another neighbouring atom at the same time it would suffer from
gigantic energetic changes which cannot occur in conserved superconducting
currents without expenses in energy. It seems unavoidable that superconducting currents are carried by closed chains of molecules (conduction chains) whose
electrons endure ongoing cyclic exchanges. Therefore, Kamerlingh Onnes compares
the closed currents in superconductors to Ampere's molecular currents.

Given our ignorance of quantum mechanics of composite systems we are far away
from being able to convert these vague ideas into a theory. We can only
connect to some questions which can be decided on experimental grounds. 
It may be seen unlikely that different atoms form conduction chains with
each other. Perhaps the transition from a superconducting metal to a different
one is never superconducting. Furthermore, it is reasonable to assume
for this reason that so far only those metals which have a relatively low melting point turned out to be superconducting; because in such [metals]
impurities are not present in the state of a real solution, but rather
in the state of small complexes which are released in the ductile state
of the metal. 
Furtheremore, there is the possibility that conduction chains cannot
carry arbitrarily small currents but only currents with a certain finite
value. This would also be accessible to experimental verification.

It is not too far stretched to assume that conduction chains can be destroyed
by magnetic fields - it is almost necessary! The same is true for
the temperature induced motion which can destroy conduction chains if
it is strong enough and if the $h \nu$ energy quanta that are being created
are big enough. This may explain the transformation of superconductors
into ordinary conductors and maybe even the sharp temperature limit of superconductors by means of an increase in temperature. Electric conductivity
at ordinary temperature may be based on ongoing thermal creation and annihilation of conduction chains. 

Phantasizing can only be excused by the momentary quandary of the theory.
It is obvious that new ways of serving the facts of
superconductivity justice have to be found. It seems probable but not
certain that the conduction at ordinary temperature is based on thermal
motion of ongoing perturbed superconductivity.
This thought is plausible considering that the frequency of the transition
of the electrons to the neighbouring atom should be closely related to the
circulation frequency of electrons in an isolated atom. One concludes
that the elementary currents of the individual conduction chains could be
significant. If this idea of elementary currents caused by quanta proves
correct it will be evident that such chains can never contain different
atoms. 

\bigskip

{\bf P.S.}

The last speculation (which by the way is not new\footnote{cf. e.g. F. Haber, Sitz.ber. Ak. Berlin, 1919, pp. 506})
is contradicted by an important experiment which was conducted
by Kamerlingh Onnes in the last couple of months. He showed
that at the interface between two superconductors (lead and tin)
no measureable Ohm resistance appears.

{\bf Acknowledgments by the translator}

It is a pleasure to thank the Kamerlingh Onnes Laboratory of the Institute of Physics
at Leiden University for kind permission to translate this article. Also, I would
like to thank Patricia T. Viele from the Edna McConnell Clark Physical Sciences Library
at Cornell University in Ithaca, New York for her support and Neil W. Ashcroft for interesting
discussions at the lunch table. Ronald Smeltzer kindly informed me of a typo in the
citation of the original article.
  
\end{document}